\documentstyle[12pt,epsfig]{article}
\begin{document}
\setlength{\parskip}{0.5cm}
\setlength{\baselineskip}{0.65cm}
\begin{titlepage}
\begin{flushright}
DO-TH 95/11 \\ RAL-TR-95-028  \\ July 1995
\end{flushright}
\vspace{0.5cm}
\begin{center}
\Large
{\bf Radiative Parton Model Analysis} \\
\vspace{0.1cm}
{\bf of Polarized Deep Inelastic Lepton Nucleon Scattering} \\
\vspace{1.5cm}
\large
M. Gl\"{u}ck, E. Reya \\
\vspace{0.5cm}
\normalsize
Universit\"{a}t Dortmund, Institut f\"{u}r Physik, \\
\vspace{0.1cm}
D-44221 Dortmund, Germany \\
\vspace{1.2cm}
\large
W. Vogelsang \\
\vspace{0.5cm}
\normalsize
Rutherford Appleton Laboratory \\
\vspace{0.1cm}
Chilton Didcot, Oxon OX11 0QX, England \\
\vspace{2.5cm}
{\bf Abstract}
\vspace{-0.3cm}
\end{center}
A leading order QCD analysis of spin asymmetries in polarized
deep inelastic lepton nucleon scattering is presented within
the framework of the radiative parton model. Two resulting sets
of plausible leading order spin dependent parton distributions
are presented, respecting the fundamental positivity constraints
down to the low resolution scale $Q^2=\mu_{LO}^2=0.23$ GeV$^2$.
The $Q^2$ dependence of the spin asymmetries $A_1^{p,n,d}(x,Q^2)$
is investigated in the range $1 \leq Q^2 \leq 20$ GeV$^2$ and shown
to be non-negligible for $x$-values relevant for the analysis
of present data and possibly forthcoming data at HERA.
\end{titlepage}
The measurements of polarized deep inelastic lepton nucleon scattering
yield direct information [1-3] on the spin-asymmetry
\begin{equation}
A_1^N (x,Q^2) \approx \frac{g_1^N (x,Q^2)}{F_1^N (x,Q^2)}
=\frac{g_1^N(x,Q^2)}{F_2^N (x,Q^2)/ \left[ 2x(1+R^N(x,Q^2)) \right] } \:\:\: ,
\end{equation}
$N=p,n$ and $d=(p+n)/2$, and $R \equiv F_L/2xF_1 =(F_2-2 x F_1)/2xF_1$,
where subdominant contributions have, as usual, been neglected.
The leading order (LO) QCD parton model relates $A_1^N(x,Q^2)$ to the
polarized $(\delta q^N,\delta\bar{q}^N)$ and unpolarized $(q^N,\bar{q}^N)$
quark distributions, where $q=u,d,s$, via
\begin{equation}
A_1^N (x,Q^2) = \frac{\sum_{q} e_q^2 \left[ \delta q^N (x,Q^2) +
\delta \bar{q}^N (x,Q^2) \right] }
{\sum_{q} e_q^2 \left[ q^N (x,Q^2) + \bar{q}^N (x,Q^2)\right] } \:\:\:
\end{equation}
due to the fact that $2g_1 = \sum_{q} e_q^2 (\delta q + \delta \bar{q})$,
$2F_1 = \sum_{q} e_q^2 (q + \bar{q})$ and $R = 0$ in LO (Callan-Gross
relation).
Henceforth we shall, as always, use the notation $\delta q^p \equiv \delta q$
and $q^p \equiv q$. The experimental results are often presented and
theoretically
analyzed in terms of $g_1^N (x,Q^2)$ extracted via the {\em measured} values of
$A_1^N (x,Q^2)$, $F_2^N (x,Q^2)$ and $R^N (x,Q^2)$ according to eq. (1).
Frequently also the assumption about the $Q^2$ independence of $A_1^N (x,Q^2)$
is
employed in regions of $x$ and $Q^2$ where it has not been experimentally
tested.
This assumption is, as is well known \cite{ans}, {\em not} theoretically
warranted since
one expects different $Q^2$-evolutions of the numerator and the denominator due
to the very different polarized and unpolarized splitting functions (except
for $\delta P_{qq}^{(0)}(x) = P_{qq}^{(0)}(x)$), respectively,
especially in the
small-$x$ region dominated by the flavor-singlet contributions.

It should be further noted that an analysis of $g_1^N (x, Q^2)$ as extracted
from
eq. (1) affords a full next-to-leading order (NLO) analysis due to the employed
$R^N (x,Q^2) \neq 0$, typically of the order of $20 - 30 \%$, and due to the
fact
that usually $2x F_1^N (x,Q^2)$ is well described only in a NLO QCD parton
calculation since this latter quantity is typically about $20\%$ smaller than
the LO
$F_2^N$. Such a NLO analysis is somewhat premature in view of the
presently scarce and inaccurate data on $g_1^N (x,Q^2)$. Some
further ambiguities supporting this attitude are the ones concerning
flavor $SU(3)$ breaking effects, mentioned below, as well as those
concerning the flavor singlet sector, in particular its gluonic
component. For all these reasons it seems appropriate to apply first
of all a LO analysis to the directly measured $A_1^N (x,Q^2)$
employing eq. (2), rather than to the derived $g_1^N (x,Q^2)$. Some
further advantages
of such an approach are that possible NLO and/or higher twist contributions
are expected to partly cancel in the ratio of structure functions appearing
in $A_1^N(x,Q^2)$, in contrast
to the situation for $g_1^N(x,Q^2)$. Therefore we shall use all presently
available data [2,4-8] in the small-$x$ region
where $Q^2 \stackrel{>}{{\scriptstyle \sim}} 1$ GeV$^2$ without bothering
about lower cuts in $Q^2$ usually introduced in
order to avoid possible nonperturbative higher twist effects as mandatory for
analyzing $g_1^N (x,Q^2)$. It should be furthermore emphasized that, within a
consistent LO analysis ($R^N = 0$, i.e. $2x F_1^N = F_2^N$), the results
cannot be expected to agree with experimental extractions of $g_1^N (x,Q^2)$
better than to about $20\%$. Alas, the extracted polarized LO parton
distributions are also reliable only to within
about $20\%$. The analysis affords some well established set of unpolarized LO
parton distributions which will be adopted from ref. \cite{glu}.

The searched for polarized LO parton distributions $\delta f(x,Q^2)$,
compatible with present data [2,4-8] on
$A_1^N (x,Q^2)$, are constrained by the positivity requirements implying
\begin{equation}
|\delta f(x,Q^2)| \leq f(x,Q^2)
\end{equation}
where $f = u, \bar{u}, d, \bar{d}, s, \bar{s}, g$, and furthermore by the sum
rules
\begin{eqnarray}
&&\Delta u + \Delta \bar{u} - \Delta d - \Delta \bar{d} =
g_A = F + D = 1.2573 \pm 0.0028 \\
&&\Delta u + \Delta \bar{u} + \Delta d + \Delta \bar{d} -
2(\Delta s + \Delta \bar{s}) = 3F -D = 0.579 \pm 0.025
\end{eqnarray}
with the first $(n=1)$ moment $\Delta f$ defined by
\begin{equation}
\Delta f(Q^2) \equiv \int_{0}^{1} dx \delta f(x,Q^2)
\end{equation}
and the values of $g_A$ and $3F-D$ taken from \cite{clo}.
It should be noted that the first moment, i.e. the total polarization of each
quark flavor $\Delta q$ and $\Delta \bar q$ is conserved, i.e.
$Q^2$-independent
as a consequence of helicity conservation at the quark-gluon vertex \cite{alt},
i.e. $\Delta P_{qq}^{(0)} \equiv \int_0^1 dx \delta P_{qq}^{(0)}(x) = 0$ and
$\Delta P_{qg}^{(0)} = 0$. The constraint equations (4) and (5) are the ones
used
in most analyses performed so far. While the validity of the Bj\o rken
$g_A$-sum-rule depends merely on the fundamental $SU(2)_f$ isospin
rotation between matrix elements of charged and neutral axial currents, the
constraint (5) depends critically on the assumed $SU(3)_f$ flavor symmetry
between hyperon decay matrix elements of the flavor changing charged weak
axial currents and the neutral ones relevant for $\Delta f(Q^2)$. We shall
refer to the results based on the constraints (4) and (5) as the $SU(3)_f$
symmetric `standard' scenario.

There are some serious objections \cite{seh} to this latter full $SU(3)_f$
symmetry. As a plausible alternative, Lipkin \cite{lip} has
suggested a `valence' scenario by assuming that the (flavor changing)
hyperon $\beta$-decay data fix only the total helicity of {\em valence} quarks
$\Delta q_V \equiv \Delta q - \Delta \bar{q}$, i.e.
\setcounter{equation}{3}
\renewcommand{\theequation}{\arabic{equation}'}
\begin{eqnarray}
\Delta u_V-\Delta d_V &=& g_A = 1.2573 \pm 0.0028 \\
\Delta u_V+\Delta d_V &=& 3F-D  = 0.579 \pm 0.025
\end{eqnarray}
by assuming only the fundamental $SU(2)_f$ isospin symmetry $u\leftrightarrow
d$. We shall alternatively implement this extreme possibility in order to
study the possible uncertainties, mentioned in the introduction,
related to the $SU(3)_f$ symmetry breaking effects relevant for the
$\Sigma^- \rightarrow n$ decay.

To understand the important (practical) difference between the above
two scenarios, let us consider the quantity
\setcounter{equation}{6}
\renewcommand{\theequation}{\arabic{equation}}
\begin{equation}
\Gamma_1^{p,n}(Q^2) \equiv \int_0^1 dx g_1^{p,n} (x,Q^2)
\end{equation}
usually extracted [2,4-8] from measurements
of $g_1^{p,n}(x,Q^2)$. In LO we have
\begin{eqnarray}
\Gamma_1^{p,n}(Q^2) &=& \pm \frac{1}{12}\Delta q_3+\frac{1}{36}\Delta
q_8+\frac{1}{9}\Delta \Sigma  \nonumber \\
&=& \pm \frac{1}{12} \Delta q_3 + \frac{5}{36}\Delta q_8+\frac{1}{3}
(\Delta s+\Delta \bar{s})
\end{eqnarray}
with the flavor nonsinglet ($\Delta q_{3,8}$) and singlet ($\Delta \Sigma$)
combinations being given by
\begin{eqnarray}
\Delta q_3 &=& \Delta u  + \Delta \bar{u}-\Delta d-\Delta \bar{d} \nonumber \\
\Delta q_8 &=& \Delta u  + \Delta \bar{u}+\Delta d+\Delta \bar{d}
-2 (\Delta s+\Delta \bar{s}) \\
\Delta \Sigma &=& \sum_q (\Delta q + \Delta \bar{q} ) = \Delta q_8+
3 (\Delta s+\Delta \bar{s}) \:\:\: . \nonumber
\end{eqnarray}
In the `standard' scenario $\Delta q_3$ and $\Delta q_8$ are entirely
fixed by eqs. (4,5) which gives for $\Gamma_1^p$, for example,
\begin{equation}
\Gamma_1^p = \frac{1}{12} (F+D)+\frac{5}{36} (3F-D)+\frac{1}{3}(\Delta
s+\Delta \bar{s}) \:\:\: .
\end{equation}
For $\Delta s=\Delta \bar{s}=0$ we recover the original estimate of
Gourdin and Ellis and Jaffe \cite{gej}, $\Gamma_{1,EJ}^p \approx 0.185$.
Therefore we need\footnote{In a LO analysis the polarized quark first
moments should be considered as effective moments since in NLO their
detailed physical interpretation depends on the convention (factorization
scheme) adopted where possibly \cite{ar} $\Delta q\rightarrow
\Delta q-(\alpha_s/4\pi)\Delta g$.} $\Delta s<0$ in order to comply
with recent experiments \cite{ada,abe1} which typically give
$\Gamma_1^p(Q^2=3$ GeV$^2)\approx 0.12 - 0.13$. In the `valence' scenario,
only the valence contribution to $\Delta q_8$ is fixed by eq. (5'), with
the entire $\Delta q_3$ still being fixed by (4') due to the assumption
$\Delta \bar{u}=\Delta\bar{d}\equiv \Delta \bar{q}$, which gives for
$\Gamma_1^p$ in eq. (8)
\begin{equation}
\Gamma_1^p = \frac{1}{12} (F+D)+\frac{5}{36} (3F-D)+\frac{1}{18}(10 \Delta
\bar{q}+\Delta s+ \Delta \bar{s}) \:\:\: .
\end{equation}
Thus, in contrast to eq. (10), a light polarized sea $\Delta \bar{q}<0$ will
account for the reduction of $\Gamma_1^p$ as required by recent experiments
\cite{ada,abe1} {\em even} for the extreme $SU(3)_f$ broken choice
$\Delta s=\Delta \bar{s}=0$! We use this latter maximally $SU(3)_f$
broken polarized sea in our 'valence' scenario, eqs. (4') and (5'),
which in addition is compatible with the $SU(3)_f$ broken unpolarized
radiative input $s(x,\mu_{LO}^2)=0$ of ref. \cite{glu}. A similar
discussion holds for $\Gamma_1^n$ and of course also for $g_1^{p,n}(x,Q^2)$.

Apart from applying the above scenarios for the polarized input
distributions to $A_1^N (x,Q^2)$ rather than to $g_1^N (x,Q^2)$,
the main ingredient of our analysis is the implementation of the
constraint equation (3) down to \cite{glu} $Q^2=\mu_{LO}^2=0.23$
GeV$^2$ which is {\em not} guaranteed in the usual studies done
so far (recently, e.g. in [16-19]) restricted to
$Q^2\geq Q_0^2=1-4$ GeV$^2$. We follow here the radiative (dynamical)
concept which resulted in the successful small-$x$ predictions of
unpolarized parton distributions as measured at HERA \cite{glu}.
A further bonus of this analysis is the possibility to study the
$Q^2$ dependence of $A_1^N(x,Q^2)$ in the small-$x$ region over a
wide range of $Q^2$ which might be also relevant for forthcoming
polarized experiments (HERMES) at HERA. The results for the
$Q^2$-dependence of $A_1^N(x,Q^2)$ are furthermore expected to hold
almost unchanged also in the full NLO analysis due to the observed
\cite{glu} perturbative stability of all the radiative model
predictions for {\em measurable} quantities such as $F_2^p(x,Q^2)$.

Turning to the determination of the polarized LO parton distributions
$\delta f(x,Q^2)$, it should be noted that $\delta g(x,Q^2)$ does
not appear explicitly in the LO expression (2) and is therefore
only weakly constrained by present data [1,2,4-8] on $A_1^N(x,Q^2)$
since it only enters indirectly via the $Q^2$-evolution  equations.
It is thus necessary to consider some reasonable constraints
concerning $\delta g(x,Q^2)$ in particular in the relevant small-$x$
region as, for example, requirements of color coherence of gluon
couplings at $x\approx 0$ (equal partition of the hadron's momentum
among its partons). This implies for the gluon and sea densities
\cite{brod}
\begin{equation}
\frac{\delta f(x,Q_0^2)}{f(x,Q_0^2)} \sim x \hspace*{1.5cm} \mbox{\rm as}
\hspace*{0.3cm} x\rightarrow 0 \:\:\: ,
\end{equation}
where the scale $Q_0$ at which this relation is supposed to hold remains
unspecified. Although not strictly compelling, eq. (12) is expected
\cite{brod} to hold at some `intrinsic' bound-state-like scale
($Q_0^2 \stackrel{<}{\scriptstyle \sim} 1$ GeV$^2$, say), but
certainly not at much larger purely perturbative scales $Q_0^2\gg
1$ GeV$^2$. Therefore we have fitted our input distributions
at $Q_0^2=\mu_{LO}^2=0.23$ GeV$^2$ using a general ansatz for the
polarized (light) sea of the form $\delta \bar{q} \sim x^{\alpha}
(1-x)^{\beta} \bar{q}$ with the result that all presently available
asymmetry data require $\alpha \approx 0.9-1.1$ for both above
scenarios. Consequently we have taken $\alpha=1$ and assumed
this power also to hold for $\delta g(x,\mu_{LO}^2)$, following
eq. (12), although present data do not significantly constrain
$\delta g$ as will be discussed below. Our optimal LO distributions
at $Q^2=\mu_{LO}^2=0.23$ GeV$^2$ subject to the above constraints
were found to be:
\begin{eqnarray}
\delta u_V(x,\mu_{LO}^2)&=& 0.718 x^{0.2} u_V(x,\mu_{LO}^2) \nonumber \\
\delta d_V(x,\mu_{LO}^2)&=& -0.728 x^{0.39} d_V(x,\mu_{LO}^2) \nonumber \\
\delta \bar{q}(x,\mu_{LO}^2) &=& -2.018 x (1-x)^{0.3} \bar{q}(x,\mu_{LO}^2)
\nonumber \\
\delta s(x,\mu_{LO}^2)&=&\delta \bar{s}(x,\mu_{LO}^2) =0.72 \delta
\bar{q}(x,\mu_{LO}^2) \nonumber \\
\delta g(x,\mu_{LO}^2) &=& 16.55 x (1-x)^{5.82} g(x,\mu_{LO}^2)
\end{eqnarray}
for the `standard' scenario (corresponding to $\chi^2=98.4/92\:$d.o.f.)
respecting eqs. (4) and (5), while
for the $SU(3)_f$ broken `valence' scenario, based on the
constraints (4') and (5'), we have:
\setcounter{equation}{12}
\renewcommand{\theequation}{\arabic{equation}'}
\begin{eqnarray}
\delta u_V(x,\mu_{LO}^2)&=& 0.726 x^{0.23} u_V(x,\mu_{LO}^2) \nonumber \\
\delta d_V(x,\mu_{LO}^2)&=& -0.668 x^{0.28} d_V(x,\mu_{LO}^2) \nonumber \\
\delta \bar{q}(x,\mu_{LO}^2) &=& -1.869 x (1-x)^{0.25} \bar{q}(x,\mu_{LO}^2)
\nonumber \\
\delta s(x,\mu_{LO}^2)&=&\delta \bar{s}(x,\mu_{LO}^2) =0 \nonumber \\
\delta g(x,\mu_{LO}^2) &=& 14 x (1-x)^{5.45} g(x,\mu_{LO}^2)
\end{eqnarray}
which corresponds to $\chi^2=96.8/92\:$d.o.f.
The unpolarized input densities $f(x,\mu_{LO}^2)$ are taken from ref.
\cite{glu} and, for obvious reasons, we have not taken into account
any $SU(2)_f$ breaking ($\delta \bar{u} \neq \delta \bar{d}$) as is
apparent from our ansatz for $\delta \bar{q}\equiv
\delta \bar{u}=\delta \bar{d}$
proportional to $\bar{q}\equiv (\bar{u}+\bar{d})/2$ which should be
considered as the reference light sea distribution for the positivity
requirement (3). The fact that $\delta s(x,\mu_{LO}^2)\neq 0$ in (13)
differs somewhat from our radiative input \cite{glu} $s(x,\mu_{LO}^2)=0$,
but for perturbatively relevant $Q^2\stackrel{>}{\scriptstyle{\sim}}
0.8$ GeV$^2$, where the leading twist-2 dominates in the small-$x$
region \cite{glu}, the positivity inequality (3) is already satisfied.
In this respect the input (13') for the `valence' scenario, with the
extreme $SU(3)_f$ breaking ansatz $\delta s(x,\mu_{LO}^2)=0$, is
more agreeable as far as our radiative (dynamical) approach is concerned.
Furthermore $|\delta q_V(x,\mu_{LO}^2)| \sim q_V(x,\mu_{LO}^2)$ as
$x\rightarrow 1$ in (13) and (13') which is also compatible with
arguments based on helicity retention properties of perturbative QCD
\cite{brod}. Finally, similarly agreeable fits to all present asymmetry
data shown below (with a total $\chi^2$ of 101 to 103 for 92 data points)
can be also obtained for a fully saturated (inequality (3)) gluon input
$\delta g(x,\mu_{LO}^2)=g(x,\mu_{LO}^2)$ as well as for the less saturated
$\delta g(x,\mu_{LO}^2)=xg(x,\mu_{LO}^2)$. A purely dynamical \cite{grvalt}
input $\delta g(x,\mu_{LO}^2)=0$ is also compatible with present
data, but such a choice seems to be unlikely in view of $\delta
\bar{q}(x,\mu_{LO}^2)\neq 0$; it furthermore results in an unphysically
steep \cite{grvalt} $\delta g(x,Q^2 > \mu_{LO}^2)$, being mainly
concentrated in the very small-$x$ region $x<0.01$, as in the
corresponding case \cite{god,grv} for the unpolarized parton distributions
in disagreement with experiment.

For calculating the evolutions of $\delta f(x,Q^2)$ to $Q^2>\mu_{LO}^2$
we have used the well known LO solutions in Mellin $n$-moment space
(see, e.g, refs. \cite{grv,gr}), with the moments of the polarized
LO splitting functions given in \cite{alt}. These were then Mellin-inverted to
Bj\o rken-$x$ space as described, for example, in \cite{grv}. We have used
$f = 3$ flavors when calculating $\delta P_{ij}^{(0)}$ and disregarded the
marginal charm contribution to $g_1^N$ stemming from the subprocess
$\gamma^{\ast}g \rightarrow c \bar{c}$ \cite{grvnp}. The LO running coupling
utilized was $\alpha_s (Q^2) = 4\pi / \beta_0 \ln (Q^2 / \Lambda^2)$ with
$\beta_0 = 11 - 2 f /3$ and $\Lambda^{(f)}$ being given by \cite{glu}
$$\Lambda^{(3,4,5)} =232, \: \: 200,\:\:  153 \:\:\: \mbox{\rm MeV} \:\:\: .$$
The number of active flavors $f$ in $\alpha_s(Q^2)$ was fixed by the
number of quarks with $m_q^2\leq Q^2$ taking $m_c=1.5$ GeV and $m_b=
4.5$ GeV.

A comparison of our results with the data on $A_1^N(x,Q^2)$ is presented
in fig.1. As already mentioned, fit results using a `saturated' gluon
$\delta g=g$ or $\delta g=xg$ at $Q^2=\mu_{LO}^2$ are very similar to
the ones shown in fig.1. Note that $A_1^N(x,Q^2)\rightarrow const.$ as
$x\rightarrow 1$. The $Q^2$-dependence of $A_1^N(x,Q^2)$ is presented
in fig.2 for some typical fixed $x$ values for $1\leq Q^2 \leq 20$ GeV$^2$.
The predicted scale-violating $Q^2$-dependence is substantial and similar
for the two rather different input scenarios (13) and (13'). In the
$(x,Q^2)$ region of present data [2,4-8], $A_1^p(x,Q^2)$ increases with
$Q^2$ for $x>0.01$. Therefore, since most present data in the small-$x$
region correspond to small values of $Q^2$ ($\stackrel{>}{{\scriptstyle \sim}}
1$ GeV$^2$), the determination of $g_1^p(x,Q^2)$ at a larger fixed
$Q^2$ (5 or 10 GeV$^2$, say) by assuming $A_1^p(x,Q^2)$ to be independent
of $Q^2$, as is commonly done [2,4-8], is misleading and might lead to
an {\em under}estimate of $g_1^p$ by as much as about $20\%$, in particular
in the small-$x$ region. The situation is opposite, although less
pronounced, for $-A_1^n(x,Q^2)$ shown in fig.2. This implies that
$|g_1^n(x,Q^2)|$ might be {\em over}estimated at larger fixed $Q^2$
by assuming $A_1^n(x,Q^2)$, as measured at small $Q^2$, to be
independent of $Q^2$. It should be emphasized that assuming $A_1(x,Q^2)$
to be {\em in}dependent of $Q^2$ {\em contradicts}, in general, perturbative
QCD as soon as gluon and sea densities become relevant, due to the very
different polarized and unpolarized splitting functions \cite{alt},
$\delta P_{ij}^{(0)}(x)$ and $P_{ij}^{(0)}(x)$, respectively (except
for $\delta P_{qq}^{(0)}=P_{qq}^{(0)}$ which dominates only in
the large-$x$ region). Moreover, the smaller $x$
the stronger becomes the dependence of the exactly calculated $A_1(x,Q^2)$
on the precise form of the input at $Q^2=\mu_{LO}^2$, as can be seen
in fig.2 for $x=10^{-3}$. For practical purposes, however, such ambiguities
are irrelevant since, according to eq. (1), $A_1(x,Q^2) \approx
2xg_1/F_2 \rightarrow 0$ as $x\rightarrow 0$ is already unmeasurably
small (of the order $10^{-3}$)
for $x\stackrel{<}{{\scriptstyle \sim}} 10^{-3}$. Thus the small-$x$ region
is not accessible experimentally for $g_1(x,Q^2)$, in contrast to the
situation for the unpolarized $F_{1,2}(x,Q^2)$. It is interesting to note
that the (approximate) asymptotic ($x\rightarrow 0$) QCD expression
\cite{bfd,cr} for $A_1(x,Q^2)$ does not even qualitatively describe
our exact LO results for $A_1(x,Q^2)$ in fig.2 for $x\geq 10^{-3}$.

In fig.3 we compare our LO results for $g_1^N(x,Q^2)$ with EMC, SMC
and SLAC-E142/E143 data as well as with the fit 'A' of ref. \cite{gs}. Despite
the fact that a LO analysis should not be expected to be more
accurate than about $20\%$, as discussed at the beginning, the EMC
\cite{ash} and E143 \cite{abe1} `data' at fixed values of $Q^2$ fall
consistently below our results in the small-$x$ region. This is
partly an artefact of the LO approximation (i.e. $R^N=0$ in
eq. (1)) and partly due to the fact that the original small-$x$ $A_1^p$-data
at small $Q^2$ have been extrapolated \cite{ash,abe1}
to a larger fixed value of $Q^2$ by
assuming $A_1^p (x,Q^2)$ to be independent of $Q^2$. According to
the increase of $A_1^p$ with $Q^2$ in fig.2, such an assumption underestimates
$g_1^p$ in the small-$x$ region at larger $Q^2$. On the contrary,
our results for $g_1^{p,d}$ do not show such a disagreement in the
small-$x$ region when compared with the SMC data \cite{ada,ade} in fig.3a
where each data point corresponds to a different value of $Q^2$ since
no attempt has been made to extrapolate $g_1^N(x,Q^2)$ to a fixed
$Q^2$ from the originally measured $A_1^N(x,Q^2)$. Our predictions
for the polarized LO parton distributions at the input scale
$Q^2=\mu_{LO}^2$ in eqs. (13) and (13') and at $Q^2=4$ GeV$^2$, as
obtained from these inputs at $Q^2=\mu_{LO}^2$ for the two scenarios
considered, are shown in figs.4a and 4b, respectively. The polarized
input densities in fig.4a are compared with our reference unpolarized LO
dynamical input densities of ref. \cite{glu} which satisfy of course the
positivity requirement (3) as is obvious from eqs. (13) and (13'). Our
resulting polarized densities at $Q^2=4$ GeV$^2$ are compared with the ones
(fit `A') of ref.
\cite{gs} in fig.4b. Since the polarized LO gluon density $\delta g(x,Q^2)$
is not strongly constrained by present experiments, we compare our
gluons at $Q^2=4$ GeV$^2$ in fig.5 with the ones which originate
from imposing extreme inputs at $Q_0^2=\mu_{LO}^2$, such as
$\delta g=g$, $\delta g=xg$ and $\delta g=0$, instead of the one
in (13') for the `valence' scenario.
The results are very similar if these extreme gluon-inputs
are taken for the `standard' scenario in (13), and the variation
of $\delta g(x,Q^2)$ allowed by present experiments is indeed
sizeable.

Finally let us turn to the first moments (total polarizations)
$\Delta f(Q^2)$ of our polarized parton distributions, as
defined in eq. (6). It should be recalled that, in contrast
to $\Delta g(Q^2)$, the first moments of (anti)quark densities
do not renormalize, i.e. are independent of $Q^2$, and thus
the first moments implemented at the input scale $Q^2=\mu_{LO}^2$
in (13) and (13') remain the same at any $Q^2$. Let us discuss
the two scenarios in turn:
\begin{description}
\item[`standard' scenario:] From the input distributions in (13) one
infers
\setcounter{equation}{13}
\renewcommand{\theequation}{\arabic{equation}}
\begin{eqnarray}
&&\Delta u_V=0.9585, \:\:\: \Delta d_V=-0.2988, \:\:\: \Delta \bar{q}
=-0.0720, \:\:\:  \Delta s=\Delta \bar{s}=-0.0519, \nonumber \\
&&\Delta g(\mu_{LO}^2)=0.444, \:\:\: \Delta g(4 \mbox{\rm GeV}^2)=1.553, \:\:\:
\Delta g(10 \mbox{\rm GeV}^2)=1.915,  \:\:\:
\end{eqnarray}
which result in $\Delta \Sigma =0.268$. This gives, using eqs. (10)
and (8),
\begin{equation}
\Gamma_1^p = 0.1506, \:\:\: \Gamma_1^n = -0.0589 \:\:\: ,
\end{equation}
which, for a LO result, is in satisfactory agreement with recent
SMC measurements \cite{ada,ade}
\begin{equation}
\Gamma_1^p(10 \mbox{\rm GeV}^2) =0.142\pm 0.008 \pm 0.011, \:\:\:
\Gamma_1^n(5 \mbox{\rm GeV}^2) =  -0.08 \pm 0.04 \pm 0.04
\end{equation}
as well as with the most recent E143 data \cite{abe2} implying
$\Gamma_1^n(2 \mbox{\rm GeV}^2)=-0.037 \pm 0.008 \pm 0.011$.
\item[`valence' scenario:] From the input distributions in (13') one
infers
\setcounter{equation}{13}
\renewcommand{\theequation}{\arabic{equation}'}
\begin{eqnarray}
&&\Delta u_V=0.9181, \:\:\: \Delta d_V=-0.3392, \:\:\: \Delta \bar{q}
=-0.0672, \:\:\:  \Delta s=\Delta \bar{s}=0, \nonumber \\
&&\Delta g(\mu_{LO}^2)=0.417, \:\:\: \Delta g(4 \mbox{\rm GeV}^2)= 1.509,
\:\:\:
\Delta g(10 \mbox{\rm GeV}^2)=1.866,  \:\:\:
\end{eqnarray}
which result in a total singlet contribution of $\Delta \Sigma =0.31$.
This gives, using eqs. (11) and (8),
\begin{equation}
\Gamma_1^p = 0.1478, \:\:\: \Gamma_1^n = -0.0617 \:\:\: ,
\end{equation}
which again compares well with the experimental values in (16).
\end{description}

In both scenarios the Bj\o rken sum rule manifestly holds due to
our constraints (4) and (4'), i.e. eq. (8) yields ($\Delta \bar{u}=
\Delta \bar{d}$)
\setcounter{equation}{16}
\renewcommand{\theequation}{\arabic{equation}}
\begin{equation}
\Gamma_1^p(Q^2) - \Gamma_1^n(Q^2) = \frac{1}{6} g_A \:\:\: .
\end{equation}
It is also interesting to observe that at our low input scale
$Q^2=\mu_{LO}^2=0.23$ GeV$^2$ the nucleon's spin is, within $20\%$,
carried by quarks and gluons, $\frac{1}{2}\Delta \Sigma +
\Delta g(\mu_{LO}^2) \approx 0.57$, according to (14) and (14'), which
implies for the helicity sum rule
\begin{equation}
\frac{1}{2}=\frac{1}{2}\Delta \Sigma+\Delta g(Q^2)+L_z(Q^2)
\end{equation}
$L_z(\mu_{LO}^2)\approx 0$. The approximate vanishing of this latter
nonperturbative angular momentum, being built up from the intrinsic
$k_T$ carried by partons, is intuitively expected for low
(bound-state-like) scales but not for $Q^2\gg \mu_{LO}^2$. From our
results for the total gluon polarization $\Delta g(Q^2)$ in (14) and (14')
it is also apparent that the $Q^2$-independent anomaly \cite{ar} contribution
$-[\alpha_s(Q^2)/4\pi]\Delta g(Q^2) \approx -0.03$ could equally
well serve$^1$ as the experimentally required negative contribution
\cite{gr1} to $\Gamma_1^N$ in (8), instead of allowing for finite
negative sea contributions $\Delta s$ and $\Delta \bar{q}$ in eqs. (10)
and (11), respectively. Such alternative factorization scheme scenarios
cannot be decided on purely theoretical grounds, but nevertheless
a consistent and convention independent analysis of the anomaly
contribution in Bj\o rken-$x$ space requires the full knowledge
of all polarized two-loop splitting functions.

Finally let us remark
that our results demonstrate the compatibility of our very
restrictive radiative model, cf. eq. (3), down to $Q^2=\mu_{LO}^2=
0.23$ GeV$^2$, with present measurements of deep inelastic spin
asymmetries. A {\sc Fortran} package containing our optimally
fitted `standard' and `valence' distributions can be obtained
by electronic mail from vogelsang@v2.rl.ac.uk .
\section*{Acknowledgement}
This work has been supported in part by the `Bundesministerium
f\"{u}r Forschung und Technologie', Bonn.
\section*{Note added}
While this manuscript was being completed, a complete calculation
of the two-loop splitting functions $\delta P_{ij}^{(1)}(x)$ in the
$\overline{{\mbox {\rm MS}}}$ factorization scheme has appeared
for the first time \cite{willi}. For the reasons stated in the
introduction we postpone a full quantitative NLO analysis to a
future separate publication.

\newpage
\section*{Figure Captions}
\begin{description}
\item[Fig.1] Comparison of our results for $A_1^N(x,Q^2)$ as obtained
from the fitted inputs at $Q^2=\mu_{LO}^2$ for the 'standard' (eq.(13))
and 'valence' (eq.(13')) scenarios with present data [2,4-8]. Our
results at different values of $x$ correspond to different values
of $Q^2$ according to experiment where each data point refers to a
different value of $Q^2$, starting at $Q^2\stackrel{>}{\scriptstyle \sim}
1$ GeV$^2$ at the lowest available $x$-bin.
\item[Fig.2] The $Q^2$ dependence of $A_1^{p,n}(x,Q^2)$ as predicted by the
LO QCD evolution at various fixed values of $x$.
\item[Fig.3a] Comparison of our 'standard' and 'valence' scenario results
with the data [2,4-6] for $g_1^{p,d}(x,Q^2)$. The SMC data correspond
to different $Q^2 \stackrel{>}{\scriptstyle \sim}1$ GeV$^2$ for $x\geq 0.005$,
as do the theoretical results; here the comparison with the GS fit
\cite{gs} is limited to $x>0.01$ since the data at lower $x$ correspond to
$Q^2<4$ GeV$^2$. The GS fit result for the E143 ($g_1^p$) data corresponds
to a fixed $Q^2=4$ GeV$^2$.
\item[Fig.3b] Same as in fig.3a but for $g_1^n(x,Q^2)$. The E142 and E143
data \cite{ant,abe2} correspond to an average $\langle Q^2 \rangle =2$
and 3 GeV$^2$, respectively.
\item[Fig.4a] Comparison of our fitted 'standard' and 'valence' input
densities in eqs. (13) and (13') with the unpolarized dynamical input
densities of ref. \cite{glu}.
\item[Fig.4b] The polarized densities at $Q^2=4$ GeV$^2$, as obtained
from the input densities at $Q^2=\mu_{LO}^2$ in fig.4a. The fitted GS
(set A) densities of ref. \cite{gs} are shown for comparison.
\item[Fig.5] The experimentally allowed range of polarized gluon densities
at $Q^2=4$ GeV$^2$ for the 'valence' scenario with differently chosen $\delta
g(x,\mu_{LO}^2)$ inputs. The 'fitted $\delta g$' curve is identical to the
one in fig.4b and corresponds to $\delta g(x,\mu_{LO}^2)$ in eq. (13'). Very
similar results are obtained if $\delta g(x,\mu_{LO}^2)$ is varied
accordingly within the 'standard' scenario.
\end{description}

\newpage

\pagestyle{empty}

\vspace*{-4.1cm}
\hspace*{-0.7cm}
\epsfig{file=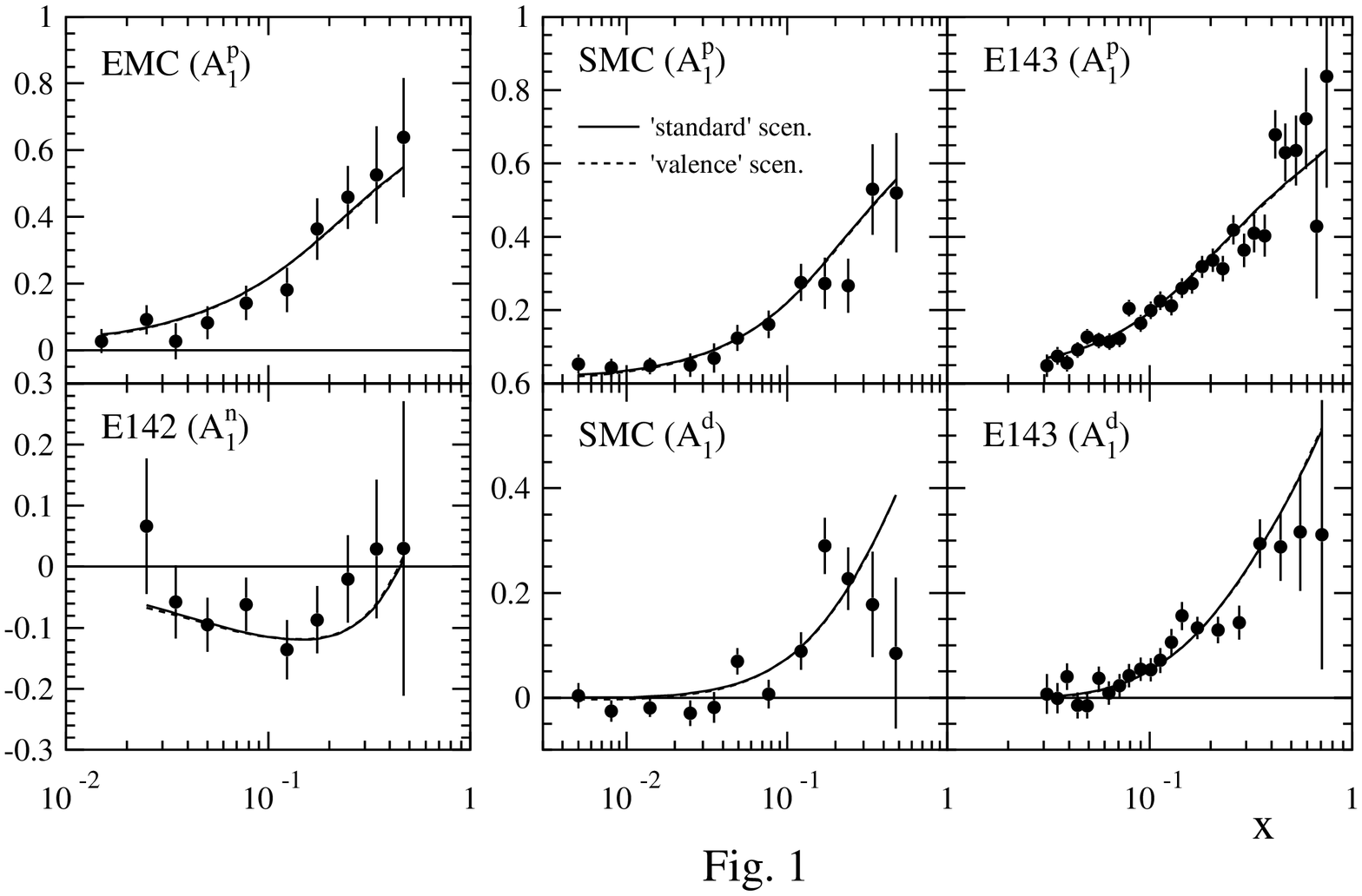,angle=90}

\newpage
\vspace*{-2cm}
\hspace*{-1.7cm}
\epsfig{file=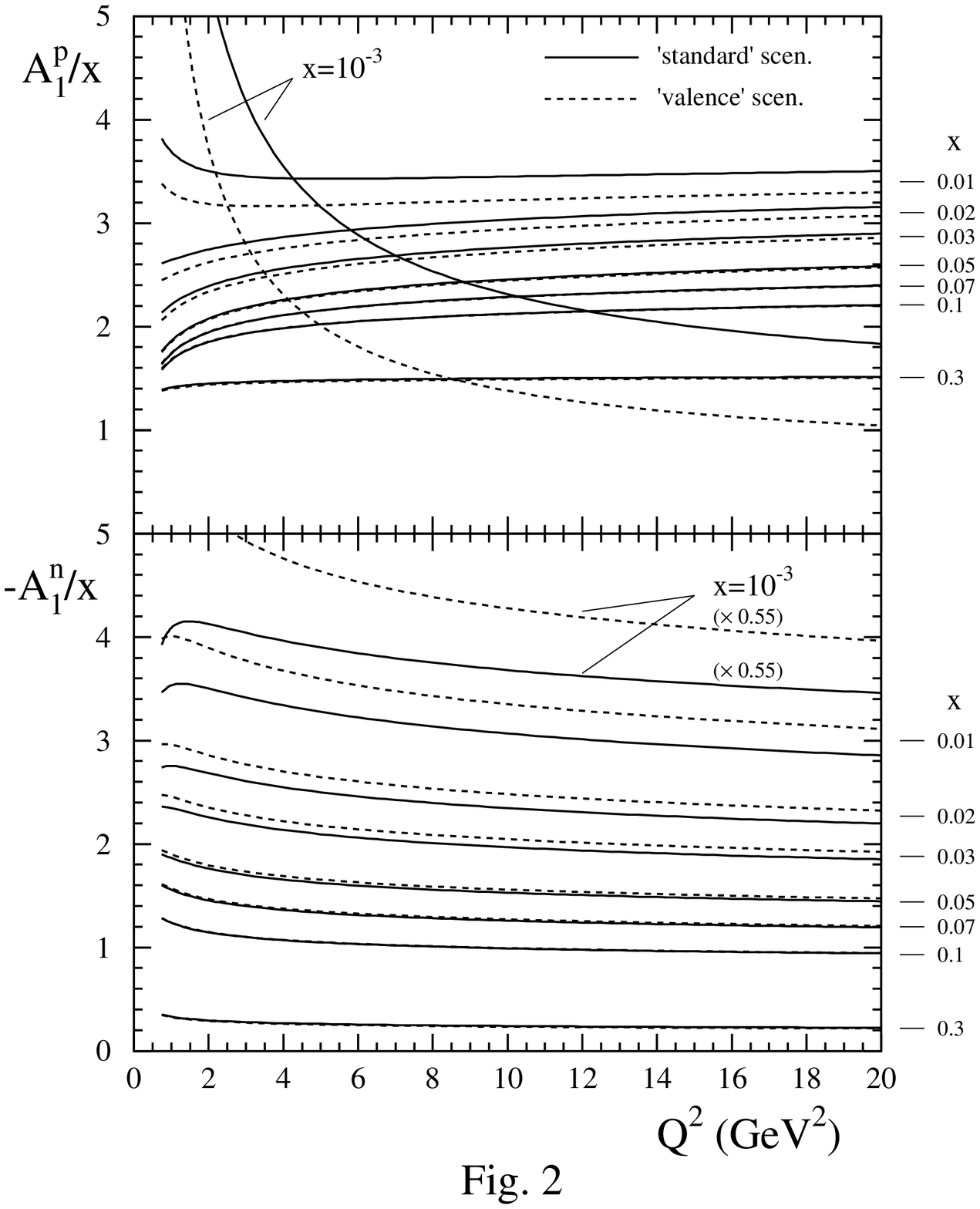}

\newpage
\vspace*{-1cm}
\hspace*{-0.7cm}
\epsfig{file=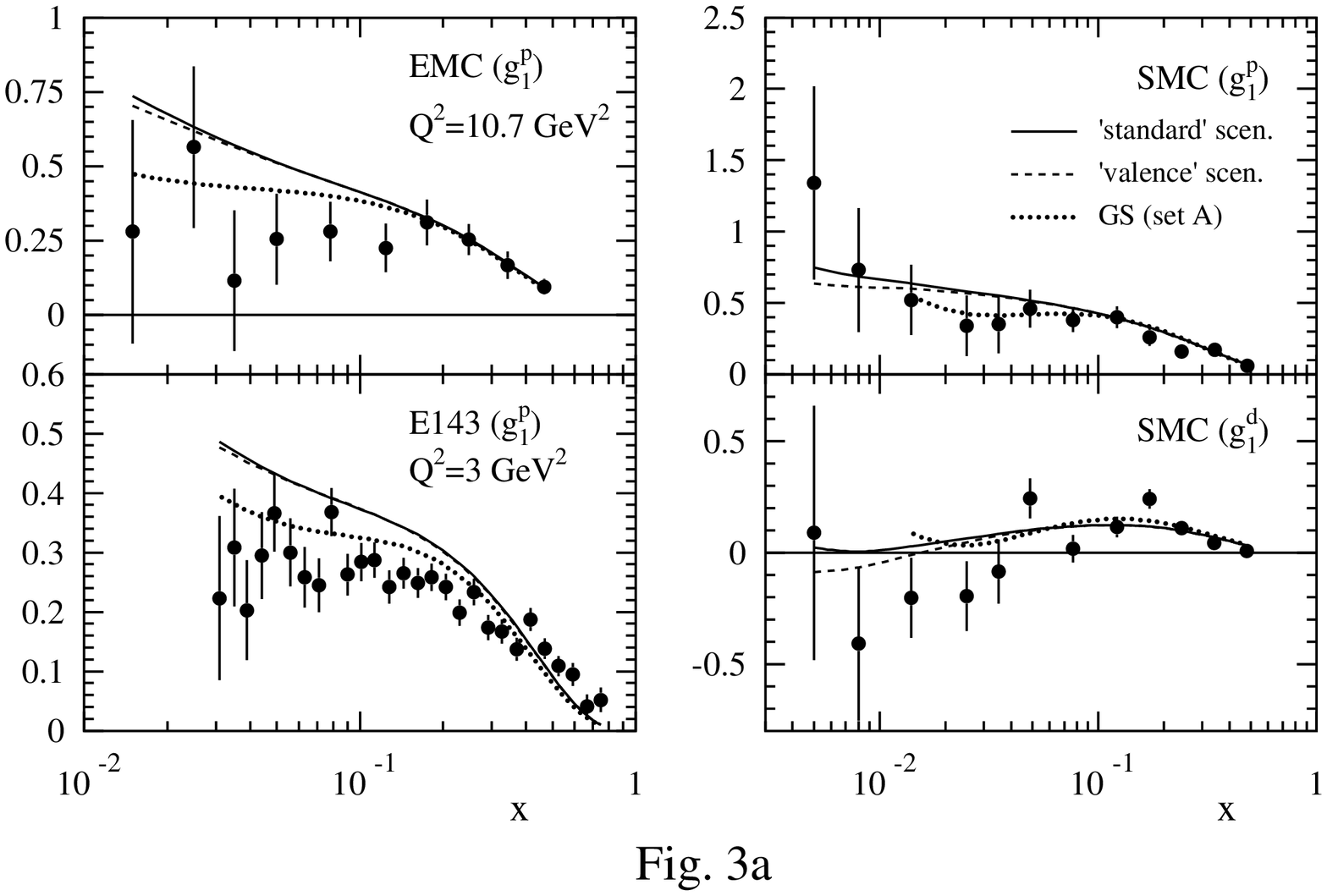,angle=90}

\newpage
\vspace*{0cm}
\hspace*{-1.3cm}
\epsfig{file=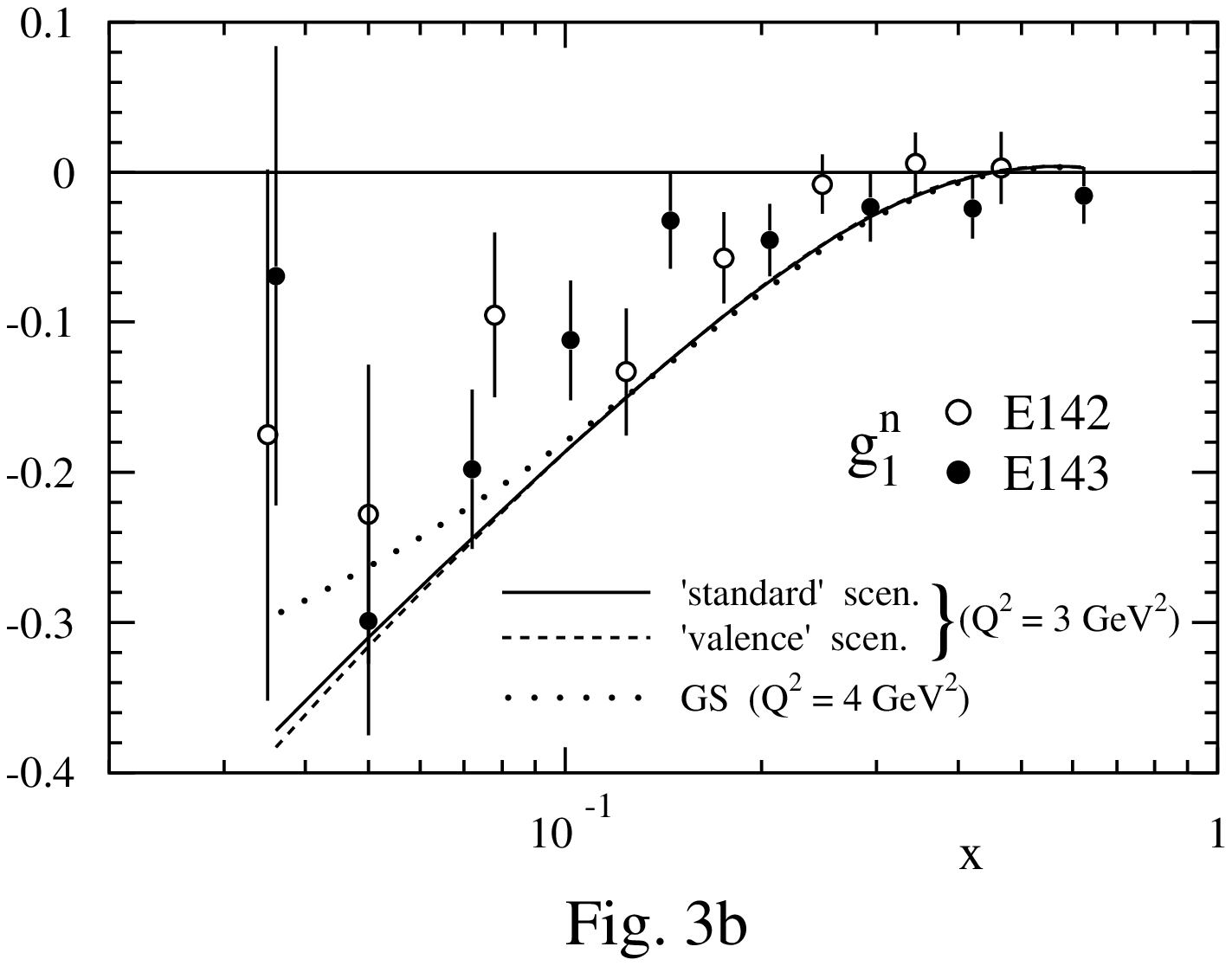,angle=90}

\newpage
\vspace*{-2cm}
\hspace*{-0.7cm}
\epsfig{file=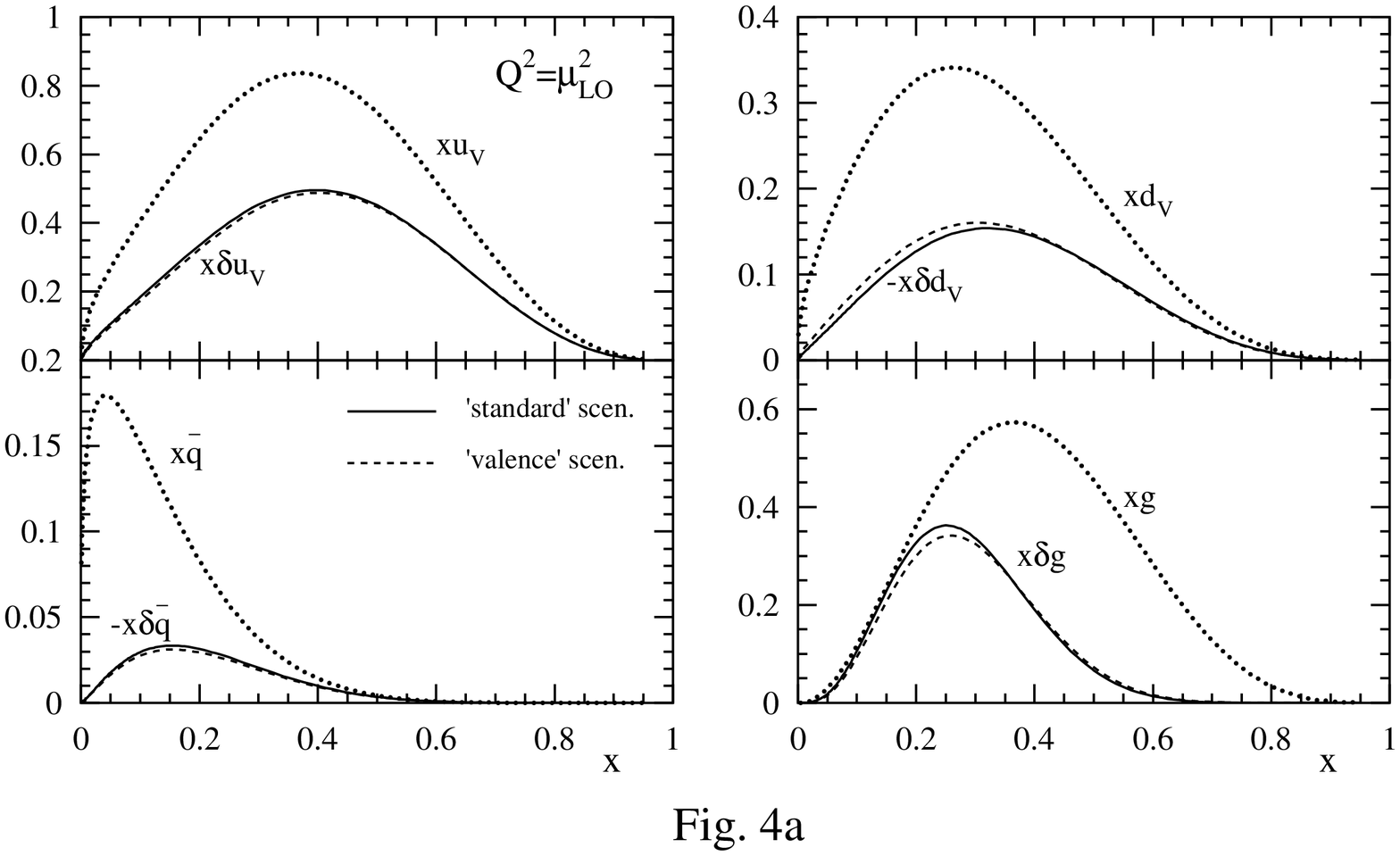,angle=90}

\newpage
\vspace*{-1cm}
\hspace*{-0.7cm}
\epsfig{file=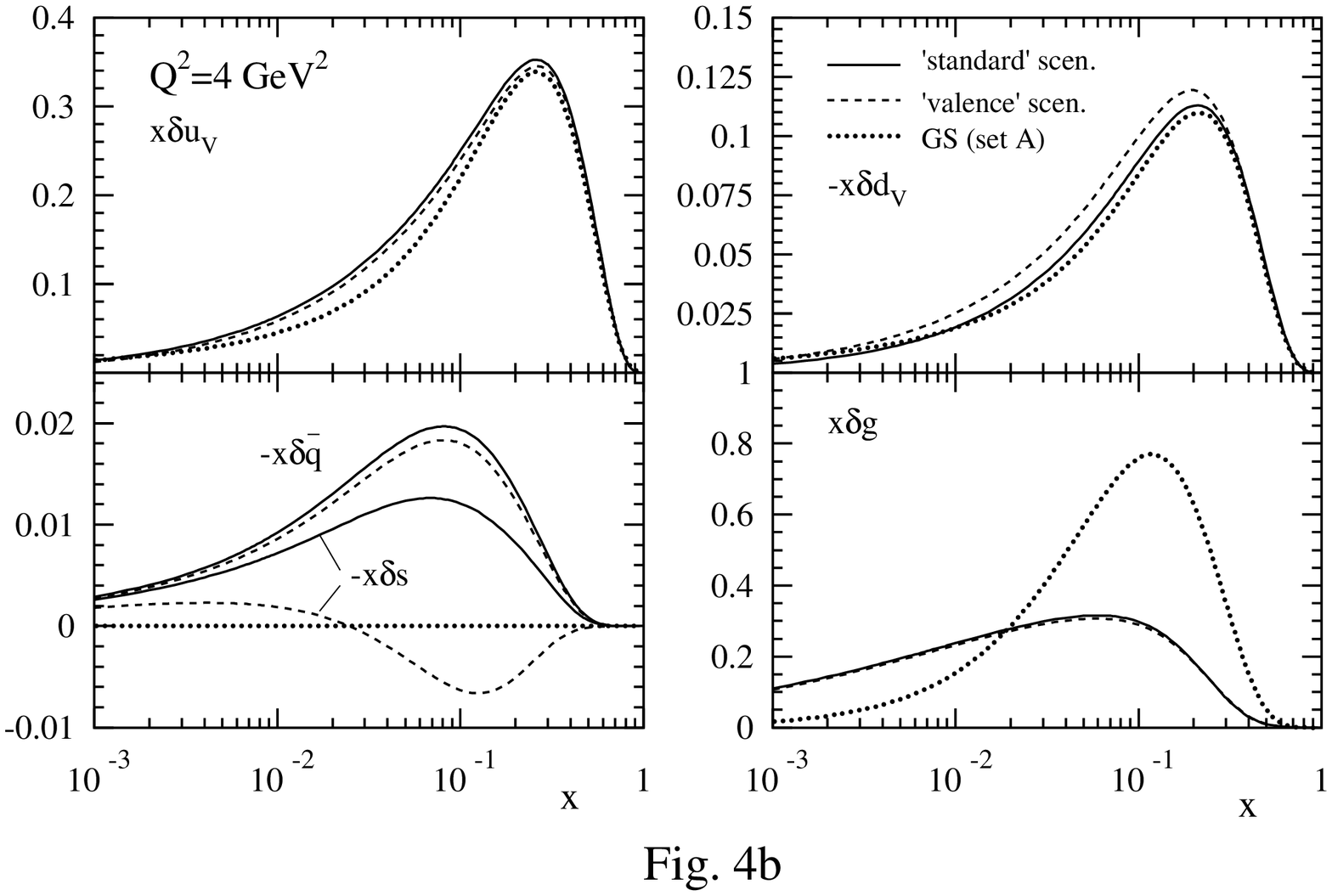,angle=90}

\newpage
\vspace*{0cm}
\hspace*{-1.4cm}
\epsfig{file=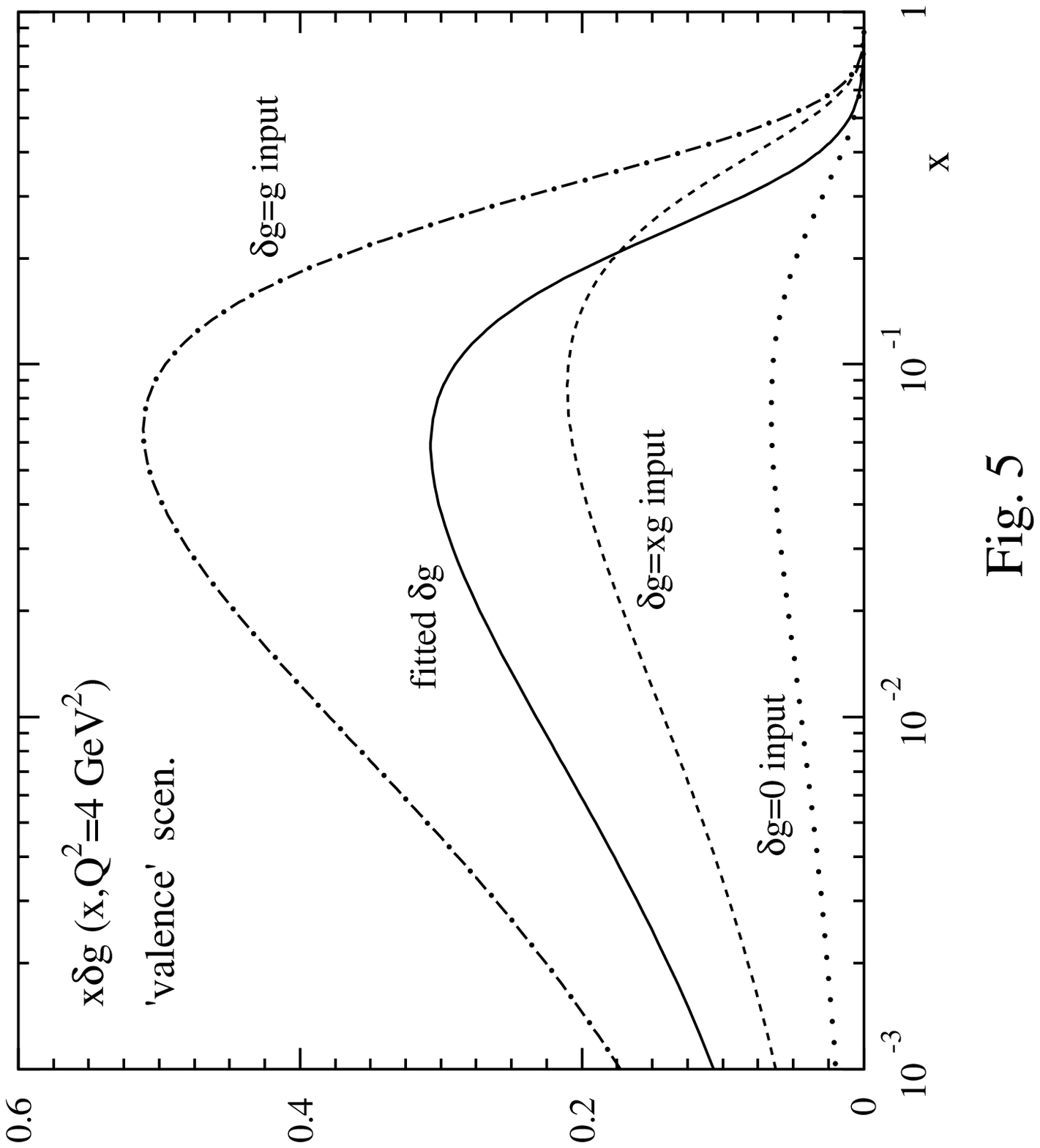,angle=90}

\end{document}